\documentclass[11pt,twoside]{article} 
 
\usepackage{asp2006}
\usepackage{graphicx} 

\begin{document}
\title{No color-morphology bimodality of AGN host galaxies}  
\author{Asmus B\"ohm\altaffilmark{1}, Lutz Wisotzki\altaffilmark{1} and the 
GEMS team
}  
\altaffiltext{1}{Astrophysikalisches Institut Potsdam, An der Sternwarte 16,
  14482 Potsdam, Germany}

\begin{abstract} 

It is still a matter of debate whether the
properties of galaxies hosting an Active Galactic Nucleus (AGN) 
are different from the properties of quiescent galaxies.
We constructed a sample of $\sim$\,50 AGN 
at a mean redshift of $\langle z \rangle \approx 0.6$
that lack a detectable optical nucleus. This characteristic
allows to study the properties of the host galaxies with much higher accuracy 
than in the case of ``normal'' AGN which show a prominent central
point source in optical images.
A comparison sample of X-ray faint, quiescent galaxies at intermediate 
redshifts shows a clear bimodality in terms of both rest-frame colors and 
morphological concentration indicators. 
In contrast to this, the AGN host galaxies comprise 
a large fraction of objects that have \emph{early-type morphologies but
relatively blue rest-frame colors}, possibly due to recent or 
ongoing star formation. 
A fraction of the ``optically dull'' AGN in our sample 
show evidence for kpc-scale absorption; 
low Supermassive Black Hole accretion rates are more likely in
other cases.
\end{abstract}

\section{Introduction}   

During the last years, growing evidence has been collected that 
Supermassive Black Holes (SMBHs) play an important role in the formation and
evolution of the galaxies they reside in.
Massive black
holes have been detected in practically all large spheroidal
galaxies that are nearby enough to resolve
their central kinematics at intrinsically small spatial scales. 
Hence, most galaxies with a spheroidal
component might undergo a phase during which the SMBH
accretes matter and is observed as an Active Galactic Nucleus (AGN). 
It is however still an open question whether the spectrophotometric and/or
morphological properties of AGN hosts
are different from those of the quiescent galaxy population.

Usually, it is a complicated matter to study galaxies
harboring an active nucleus due to the bright central point source.
A very good modelling of the Point Spread Function is crucial to
analyse the light profile of the host galaxy, in particular at
intermediate and high redshifts \citep[e.g.][]{San04}.
Here, we will utilise a sample of AGN \emph{without a detectable optical nucleus}
to study the host colors and morphologies much more robustly than it would be
feasible for ``optically normal'' AGN.

\section{Sample Selection}

\begin{figure}[t]
\includegraphics[width=6.65cm,angle=270,bb=53 38 555 533]{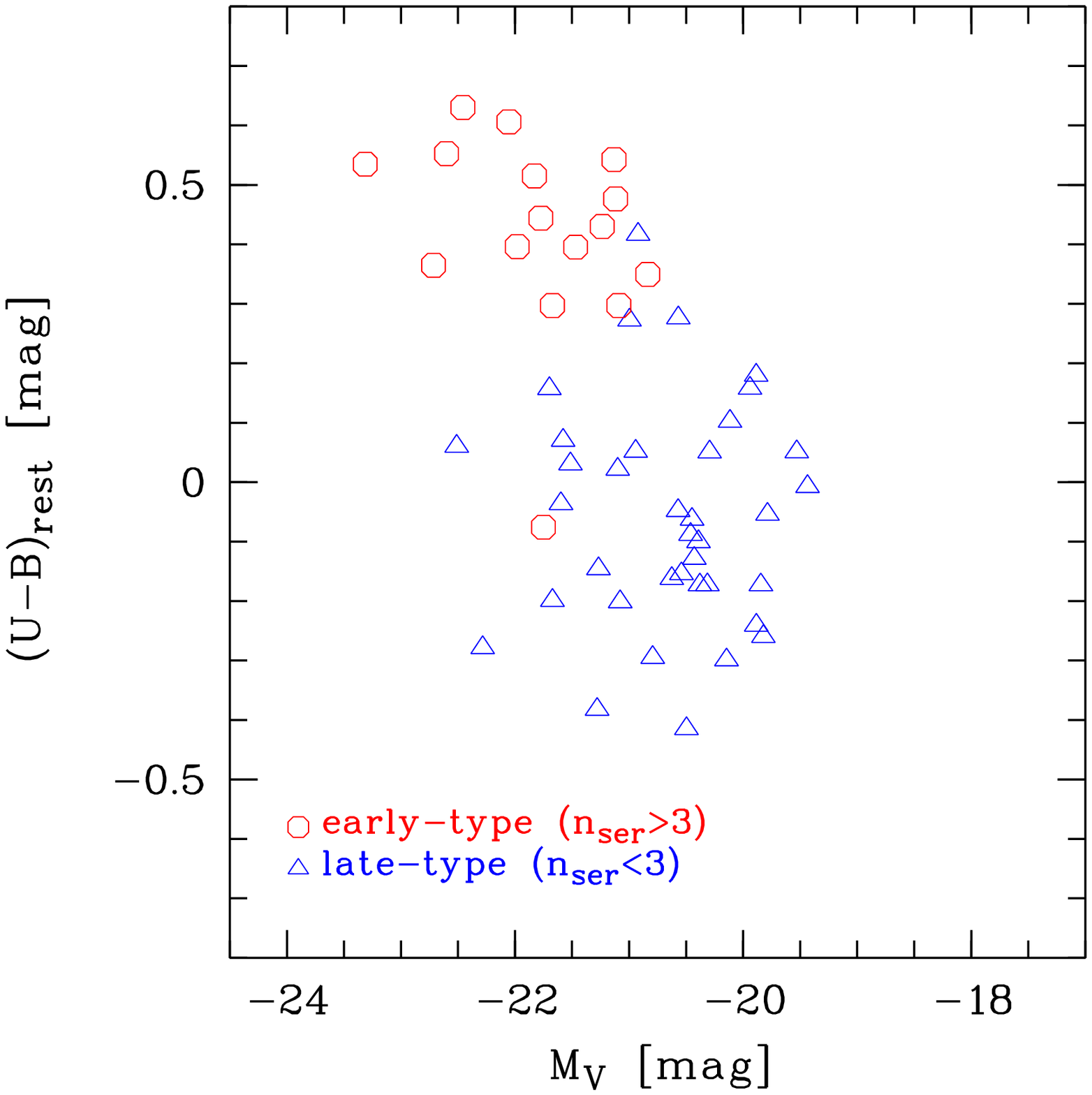}
\hfill
\includegraphics[width=6.65cm,angle=270,bb=53 38 555 533]{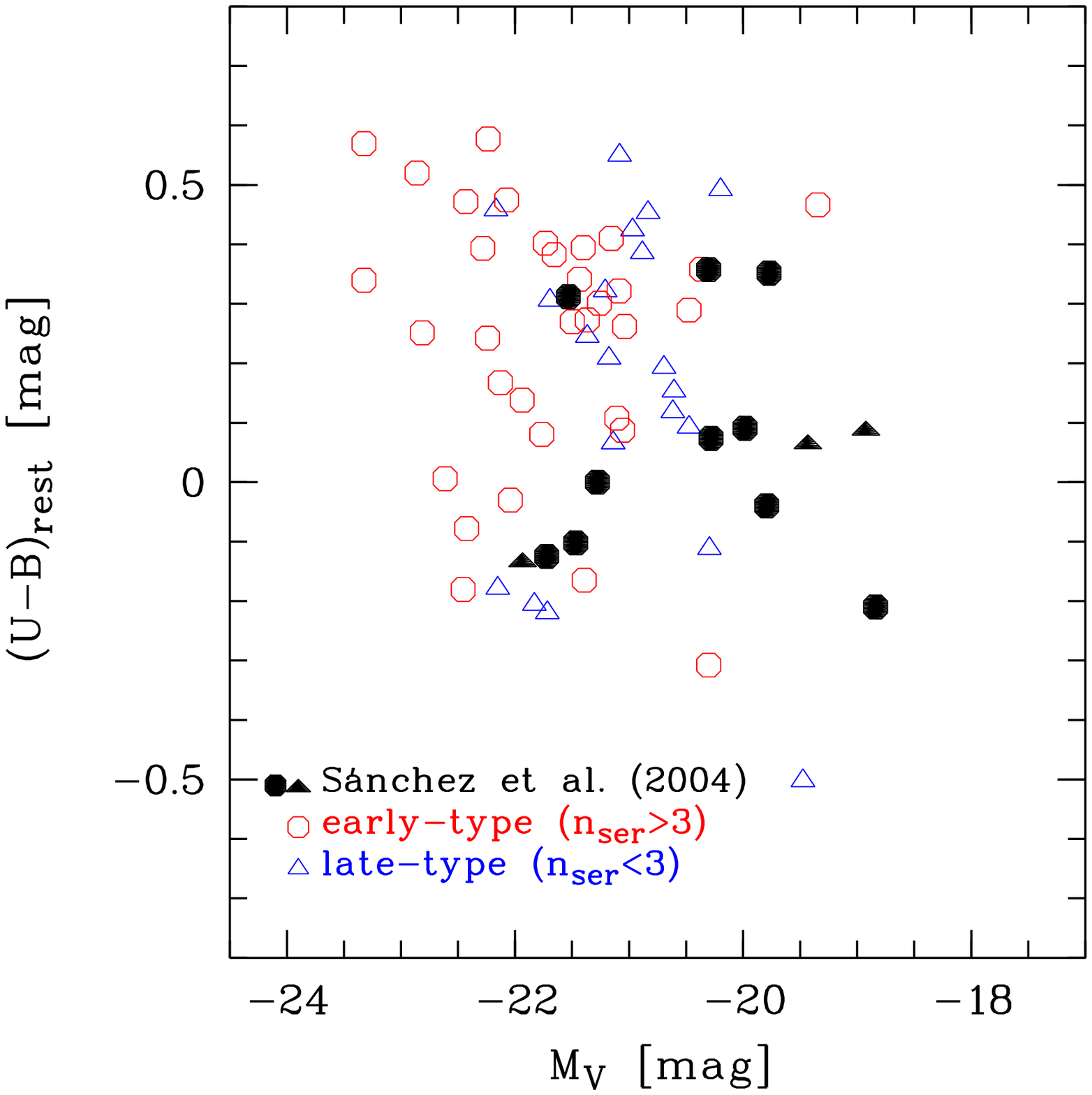}
\caption{\emph{left:} color-magnitude diagram of quiescent galaxies at redshifts
  $\langle z \rangle \approx 0.6$.  Early- and late-type morphologies are indicated
  by circles and triangles, respectively.
\emph{right:} our sample of AGN at the same redshifts (open symbols). 
For comparison, the type-1 AGN (with prominent optical
  nuclei) from \citet{San04} are depicted by solid symbols.
A flat cosmology with $\Omega_\lambda=0.7$, $\Omega_m=0.3$ and
$H_0=70$\,km\,s$^{-1}$\,Mpc$^{-1}$ has been assumed.
}
\end{figure}

For the construction of our sample, we combined
deep ground-based multi-color photometry from COMBO-17 \citep[][]{Wol03},
HST/ACS imaging from the GEMS survey \citep[][]{Rix04} and the X-ray point
source catalogue of the (Extended) Chandra Deep Field South
\citep[CDFS,][]{Gia02,Leh05}.
We pre-selected X-ray detections with unambiguous optical counterparts and 
X-ray luminosities that indicate an accreting SMBH.
All objects classified as broad-line (type-1) AGN either in the 
COMBO-17 photo-z catalogue or on the basis of optical spectra from
\citet{Szo04} were \emph{rejected}.
Using the GALFIT package by \citet{Pen02}, we also performed 
2-D light profile fits
and tossed all galaxies from the data set 
that showed even only a \emph{potential} central point source ($\ge$\,10\,\% of the
total flux) in the ACS F606 or F850 images.

The remaining sample holds 53 objects in the range $0.3<z<1.2$ with a
median of $\langle z \rangle \approx 0.6$. 
More than 90\% of the optical spectra either show emission lines that can be 
attributed entirely to star formation, or no emission lines at all
\citep[``optically dull'' AGN, e.g.][]{Mor02}. 
However, for the majority of the objects, no spectra are available, hence a
fraction of these might be narrow-line AGN.
As a quiescent comparison sample, we selected COMBO-17 galaxies at the same
redshifts that were not detected in X-rays. Note that 
exactly the same morphological de-selection criteria as for the AGN were 
applied to these objects.

\section{Results and Discussion}

Fig.~1 shows the color-magnitude diagrams of the quiescent galaxies (left
panel) and the AGN hosts (right panel).
In the case of the former, 
the ``blue cloud'' and a ``red regime'' 
are well separated. Moreover, the objects are equally separated by
morphological type (objects with a S\'ersic index 
$n_{\rm ser}>3$ are classified as early-types, those with $n_{\rm ser}<3$ as 
late-types). In contrast, the AGN sample comprises
early-type galaxies with relatively blue colors as well as late-types with
relatively red colors~-- the AGN hosts do not populate two distinct regimes
in color-magnitude space.
A similar result has been found by \citet{San04} for a smaller sample of
type-1 AGN. However, the presence of a bright optical nucleus in
these objects~-- which had to be fitted simultaneously with the host profile~--
did not allow to robustly determine the S\'ersic index but only to
roughly characterise the host galaxies as either ``disk-like'' or ``bulge-like''.

\begin{figure}[t]
\hspace*{1.3cm}
\includegraphics[width=7cm,angle=270,bb=56 37 550 772]{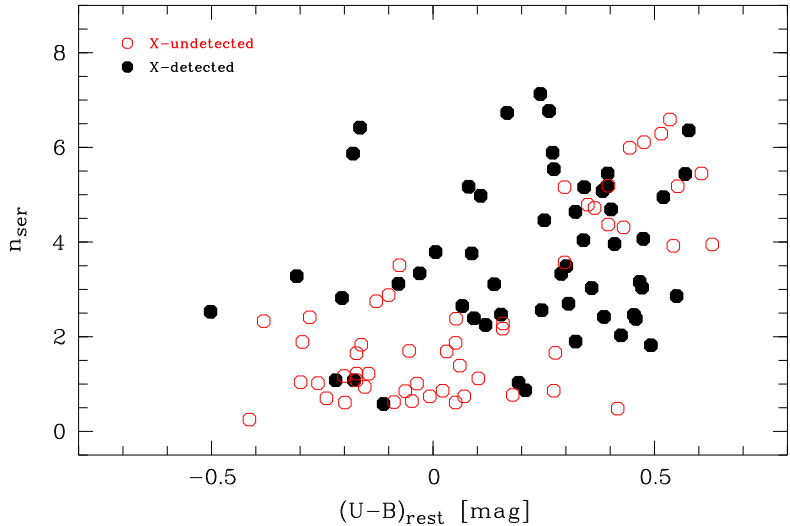}
\caption{S\'ersic index versus rest-frame $U-B$ color for the quiescent galaxy
  sample (open circles) and the AGN hosts (solid circles).
}
\end{figure}

In terms of the surface brightness 
profile analysis, the absence of a nuclear point source
in the optical images of our X-ray selected sample is a big advantage. 
Fig.~2 shows
the S\'ersic indices as a function of rest-frame $U-B$ color. The quiescent,
X-ray-undectected sample (open symbols) shows a clear bimodality comprising
blue galaxies with $n_{\rm ser} \approx 1-2$ (disk-dominated)
and red galaxies with $n_{\rm ser} \ge 4$ (bulge-dominated).
The bimodality is totally absent in the case of the ``optically dull''~/~narrow-line
AGN sample. This striking difference is probably a combination of three
aspects: firstly, 
pure disk systems ($n_{\rm ser} \approx 1$) are scarce in the AGN
sample, which is a natural consequence of the SMBH mass-bulge mass correlation
\citep[e.g.][]{Geb00,Fer00}.
Secondly, a significant fraction of the early-type AGN hosts show
bluer colors than their quiescent counterparts, possibly due to recent or
ongoing star formation.
Thirdly, many hosts have intermediate S\'ersic indices indicating a 
disk$+$bulge morphology and relatively red colors. A closer inspection
reveals that the majority of these systems are observed edge-on, i.e.~the red
colors probably arise from intrinsic absorption in the plane of the disk component.

This interpretation is  supported by Fig.~3, where the X-ray hardness
ratio~-- defined as $(H+S)/(H-S)$, where $S$ and $H$ are the resp.~fluxes in the
0.5\,\dots2\,keV and 2\,\dots10\,keV bands; 
large values of the hardness ratios indicate
absorption~-- is shown as a function of X-ray luminosity. Most of the sources
in the CDFS \& E-CDFS
at similar redshifts are distributed around a hardness ratio of 
approx.~$-0.5$, these are type-1 (broad-line) AGN. 
The majority of the objects in our sample show evidence for
absorption, in particular those which have small axis ratios $b/a<0.5$,
hence are observed edge-on. 
The phenomenon of ``optically dull'' AGN could therefore partly
be caused by absorbing material that is surrounding the nucleus out to scales 
of a kpc, thereby blocking the Narrow Line Region emission (Rigby et al.~2006 came
to a similar conclusion using a smaller sample).
However, Fig.~3 also shows that a fraction of the AGN hosts in our sample
are observed relatively face-on and have moderate hardness ratios. Since the
X-ray luminosities are also relatively low in most
of these cases, the lack of an optical nucleus might be due to a relatively 
low black hole
mass, a low accretion rate or even a truncated inner accretion disk.

\begin{figure}[t]
\hspace*{1.3cm}
\includegraphics[width=7cm,angle=270,bb=56 40 553 770]{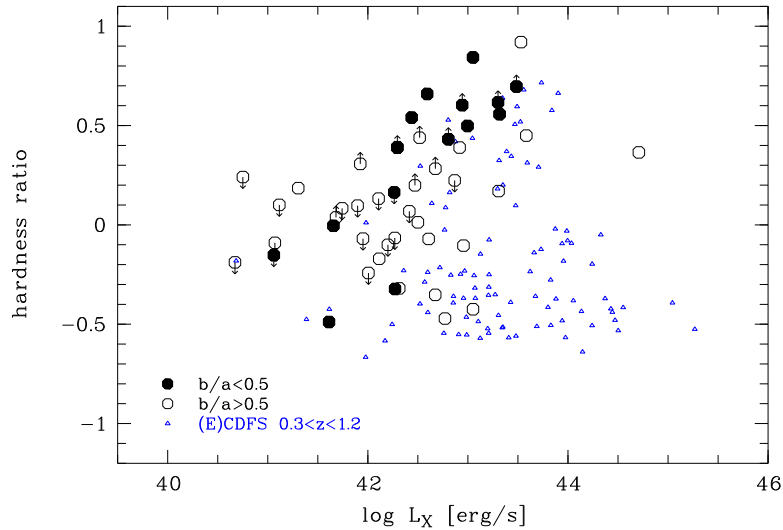}
\caption{The X-ray hardness ratio as a function of X-ray luminosity. The large
  circles denote our X-ray selected AGN sample
(solid and open circles represent edge-on and face-on galaxies,
respectively), the small triangles show
  the other X-ray sources in the (Extended) Chandra Deep Field South at similar
  redshifts. Arrows indicate upper or lower limits on the hardness ratio.
}
\end{figure}

\acknowledgements 
This work was funded by the ``Deutsches Zentrum f\"ur Luft- und Raumfahrt'' 
(50\,OR\,0404).

\end{document}